# A simple synthesis method for growing single crystals of a copper coordination polymer [Cu(C$_2$O$_4$)(4-aminopyridine)$_2$(H$_2$O)]$_n$, and its theoretical and physical properties studies

*George Mathew,[†] Sebastian Francis,[‡] Neeraj K. Rajak,[†] Praveen S. G.,[†] C. V. Tomy,[¶] and D. Jaiswal-Nagar*[†]*

[†]*School of Physics, IISER Thiruvananthapuram, Vithura, Thiruvananthapuram-695551, India*

[‡]*School of Chemistry, IISER Thiruvananthapuram, Vithura, Thiruvananthapuram-695551, India*

[¶]*Department of Physics, IIT Bombay, Powai, Mumbai-400076, India*

E-mail: deepshikha@iisertvm.ac.in

## Abstract

This work reports on a novel and simple synthetic route for the growth of metal-organic crystal [Cu(C$_2$O$_4$)(4-aminopyridine)$_2$(H$_2$O)]$_n$ of large size using the technique of liquid-liquid diffusion or layer diffusion. Single crystal X-ray diffraction measurements revealed a very good quality of the grown single crystals with a small value 1.101 of goodness of fit R. Rietveld refinement done on powder X-ray diffractogram obtained on few single crystals crushed together revealed a very small value of R as 3.45, indicating very good crystal quality in a batch of crystals. Density functional theory with three different basis sets generated the optimized geometry of a monomeric unit as well as its vibrational spectra.

Comparison between experimentally obtained bond lengths, bond angles, IR frequencies etc. suggest (B3LYP/LanL2DZ, B3LYP/6-311++ G(d,p) basis set to describe the properties the best. Magnetic susceptibility measurements confirm the metal-organic crystal [Cu(C$_2$O$_4$)(4-aminopyridine)$_2$(H$_2$O)]$_n$ to be a very good representation of a spin 1/2 Heisenberg antiferromagnet.

## 1. Introduction

Metal-organic coordination polymers are coordination network with organic ligands that link multiple metal centers together, in an infinite array. These polymers have attracted much interest in material science due to their diverse applications in luminescence,[1-3] magnetism,[4-10] gas adsorption,[11-17] and catalysis.[18-21] Low-dimensional magnetism exhibited by such materials is an active research area because of their potential applications as one-dimensional molecular magnetic nanowires and magnetic storage devices.[22] These polymers provide an excellent platform for realizing low dimensional magnets because of the possibility of tuning both the bridging ligands as well as the co-ligands, thereby, controlling the extent of magnetic exchange interactions and dimensionality.[23] Short bridges such as oxo-, cyano- or azido-ligands are used as bridging ligands to achieve moderately strong coupling between spin centers.[24-26] An excellent example of a quasi-one-dimensional magnet is [Cu(C$_2$O$_4$)(4aminopyridine)$_2$(H$_2$O)]$_n$ (abbreviated to **1** henceforth), where oxalate molecules are the bridging ligand that links magnetic centers together, and forms a remarkable physical realization of a spin 1/2 antiferromagnetic Heisenberg chain[27-29] (AfHc), whose excitation spectrum can be investigated using inelastic neutron scattering measurements.

However, since the flux of neutrons in a typical time-of-flight neutron scattering measurement is weak, a large mass and size of the metal-organic crystal is warranted in order to obtain a good signal.[30] Amongst the several approaches to overcome these limitations, the simplest one is to increase the diffracting volume of the crystals.[31,32] Single crystals of the antiferromagnetic Heisenberg chain coordination polymer **1** were first synthesized by Castillo *et al.*,[27] and later by Prokofiev *et al.*,[28] using the slow diffusion technique. The crystals grown using Castillo's method were very small that grew as dendrites. Because of their small size, these crystals may not be suitable for neutron scattering experiments. Prokofiev's technique resulted in large-sized crystals (~ 1 mm in size). However, the pH of the starting reactants had to be critically controlled, else precipitation of the reactants itself took place. In this paper, we report on a novel and simple synthetic route to grow large-sized single crystals of **1** using the layering technique in which one solution is layered on top of another such that the precipitation process is slowed down, to ensure large-sized crystals. Using this technique, we have grown crystals that are not dendritic in nature and have a larger volume, making it an appealing candidate for various neutron scattering experiments. The crystals were then characterized with various experimental techniques. Density functional theory with three different basis set was employed to generate the monomeric unit of **1** and comparison with experimentally obtained bond lengths, bond angles and vibrational spectra made.

**Experimental Section**

**2.1. Materials and General Methods**

All the reagents for synthesis were obtained commercially and used as received; 4-aminopyridine (Merck-Aldrich, 99%), Potassium bis(oxalato)cuprate(II)dihydrate (Merck-Aldrich, 99.999%), Isopropanol (Merck-Aldrich, 99.5%). Fourier transform infrared (FTIR) measurements were done using Shimadzu's spectrometer (Model No. IRPrestige-21) in the 4000-450 $cm^{-1}$ range using KBr pellet. Thermogravimetric analysis (TGA) and differential scanning calorimetric analysis (DSC) measurements were done in TA Instruments thermal analyzer (Model no. SDT Q600) till 1000 °C at a slow heating rate of 5 °C/min. under nitrogen environment. Magnetization measurements in the temperature range of 1.9 K to 300 K were done on a vibrating sample magnetometer (VSM) mounted on a Physical Property Measurement System (PPMS) from Quantum Design (Model No. Evercool-II). Lower temperature measurements in the range of 0.49 K to 2 K were done on a Helium 3 insert attached to Quantum Design's Superconducting Quantum Interference Device (SQUID) magnetometer (Model iHelium3). In all the experiments, the crystals were cooled in the zero field (ZFC cooling) mode to the lowest possible temperature after which a field of 1 T was applied and the data collected in the warming up mode.

## 2.2. Synthesis

In the present study, crystal growth was based on a binary layering technique (layer diffusion) that involves slow diffusion of one solvent into another resulting in the formation of crystals at the liquid/liquid boundary (**Figure 1**). As mentioned in the introduction, **1** consists of a bridging ligand and a side-ligand which are oxalate molecule and 4-aminopyridine respectively. The source of oxalate molecule is potassium bis(oxalato) cuprate(II) dihydrate. Choosing the most suitable solvent is the main parameter deciding the type, morphology and structure of the resultant crystal. A solvent in which the potassium bis(oxalato) cuprate(II) dihydrate was moderately soluble was chosen as the first solvent, which in this case, was distilled water. A solvent which was comparatively less dense than the first one and within which 4-Aminopyridine was sparingly soluble formed the second solvent, which was isopropanol. Details of the crystal growth is described in the supplementary. We obtained dark blue crystals with 65-70% yield (based on copper), as shown in the inset (a) of Figure 1. The formed crystals were filtered out and rinsed with a minimal amount of cold distilled water. The remaining excess solvent was removed by blotting the crystals on a piece of clean tissue paper. Inset (b) of Figure 1 shows images of a few extracted crystals that are at least 1 mm x 1 mm x 0.4 mm in size, similar to those obtained by Prokofiev et al. [28]

## 2.3. X-ray data collection and structure determination

A tiny crystal was subjected to single crystal x-ray diffraction (SCXRD) on a Bruker Kappa APEX II CCD diffractometer by the $\omega$ scan technique using graphite monochromatized MoK$_\alpha$ radiation ($\lambda$ = 0.71073 Å) at room temperature (293 K). A prismatic crystal with dimensions 0.200 x 0.200 x 0.100 mm$^3$ was carefully selected under a microscope and mounted on a glass fiber. Accurate unit cell parameters and orientation matrix were determined by least-squares treatment of the setting angles of 5402 reflections, of which 1382 reflections were independent, in the $3 \leq 2\theta \leq 26°$ range. The minimum and maximum normalized transmission factors were 0.736 and 0.854. Atomic positions were located by Direct Methods with the structure solution program SHELXT and were then refined by full-matrix least-squares calculations based on F$^2$ using the program SHELXL. All non-hydrogen atoms were refined anisotropically. The positions of hydrogen atoms were added in idealized geometrical positions. Details of the crystallographic data as well as structural refinement parameters for **1** are given in **Table 1**. Details of bond lengths, angles and dihedral angles are given as supplementary information. The value of goodness-of-fit R, 1.101, indicates the quality of the crystals.

## 2.4. Computational Details

The quantum chemical calculations were done using density functional theory (DFT) method using the Gaussian '16 program package. [33] For calculations of the molecular geometry, atomic co-ordinates from the crystal structure obtained in section 2.3, were input into the Gaussian software. Geometry optimization, vibration frequency calculations and IR spectrum calculations of **1** have been carried out by Lee Yang Parr functional correlation (B3LYP). Three different kinds of basis sets are commonly used in calculations for systems having metal centres and organic parts: (i) 6-311G++(d,p), (ii) LanL2DZ and (iii) An ONIOM model with LanL2DZ for the Cu atom (metal centre) and 6-311G++(d,p) for the rest of the atoms (organic part).[34,35] The charge and spin multiplicity in all quantum chemical computations were taken as 0 (neutral) and 2 (doublet) respectively. The computed vibrational modes, obtained using Gaussview 6 program, [36] have been used for the molecular structural and vibrational analysis. The calculations converged to an optimized geometry since there were only real harmonic vibrational wavenumbers, revealing the localization of energy minima. Harmonic infra-red vibrational wavenumbers were calculated for the fully optimized molecular geometry. [34,37] The differences between calculations and experiments are accounted for by scaling the generated vibrational frequencies using standard scaling factor 0.9679. [38]

## 2. Results and discussion

### 3.1. Crystal structure details

The crystal structure of **1** consists of $Cu^{2+}$ (spin-1/2) ions, bridged by oxalate molecules ($C_2O_4$), forming chains along the crystallographic c-axis, as shown in **Figure 2 (a)**. Aminopyridine and water molecules form side-ligands, respectively. The polymer chains lie at a distance of 6.36

Å in the b-direction, as observed in Figure 2 (b), with the interlayer space being filled by water molecules. The spin chains are well separated (8.32 Å) and magnetically screened by non-magnetic amino-pyridine groups along the a-axis. The screening of the $Cu^{2+}$ spins by water molecules along the b-direction and amino-pyridine groups along the a-direction makes it a quasi-one-dimensional system along the c-axis. Copper has a coordination of a square pyramid with Cu lying within the basal plane formed by oxygen atoms of two oxalate ligands and the nitrogen atoms of two 4-aminopyridine ligands. The bond distances are 2.006 Å and 1.998 Å for the basal Cu-O and Cu-N bonds, respectively. The basal plane of the square pyramidal structure is parallel to the ac plane. The apical Cu-O bond (2.290 Å) connects the copper atom with the oxygen of the water molecule. [27,28]

Figure 2 (c) shows the geometry of the oxalate ligand that acts as a bridge between two neighboring copper ions. The oxalate ligand, is not planar and the two $CO_2$ entities (O2-C6-O1 and O4-C7-O3) are twisted ~ 28° with respect to each other around the C6-C7 bond. The values of the dihedral angles (O1-C6-C7-O3) and (O2-C6-C7-O4) are 28.4° and 27.4° respectively. The distance between the two oxygen atoms O6 and O8 is 2.788Å. The oxalate binds the copper ions only through two oxygen atoms belonging to the two carboxylate moieties (O6, O8) of the oxalate. The remaining two oxygen atoms are free and do not bind to the copper ions. This situation is very rare and is usually not found where typically both the oxygen atoms of the oxalate molecules bind to the metal ions. [39,40] The free oxygen atoms of the carboxylate moieties (O7, O2) make an extensive network of hydrogen bonds with water molecules shown as dashed lines in Figure 2 (b) and (c).

The coordination sphere of copper(II) governs the electronic structure of the $3d^9$ Cu(II) ion and the nature of the orbital describing the unpaired electron ("magnetic orbital"). From the short metal-ligands distances (around 2Å), it can be deduced that the magnetic orbital is of the $d_{x^2-y^2}$ type with x roughly along O6-Cu-O1 and y along N2-Cu-N4. [41] A schematic representation of the exchange pathway mediated between the two neighboring copper ions through the oxygen atoms of the bridging oxalate molecule is shown in Figure 2 (d). The twist of the oxalate ligand can be clearly seen in Figure 2 (d). The twist decreases the overlap of magnetic orbitals and can partly explain the low value of the coupling constant $J/k_B$ in **1**.

The asymmetric unit of **1** consists of a Cu atom attached to a 4-aminopyridine molecule, three oxygen atoms, one carbon atom and one hydrogen atom as shown in **Figure 3 (a).** The simplest unit obtained by optimizing the geometry of **1** that is generated by using the density functional theory (i) B3LYP/6-311++G(d,p) basis set, (ii) B3LYP/LanL2DZ basis set and (iii) a combination of B3LYP/6-311++G(d,p), B3LYP/LanL2DZ (ONIOM) basis sets as described in the computational details section, are presented in Figs. 3 (b)-(d) respectively. From the Figure, it can be clearly seen that the simplest unit generated employing (i) B3LYP/6-311++G(d,p) basis set does not match with the asymmetric unit generated from SCXRD. In comparison, the matching between (ii) B3LYP/LanL2DZ basis set and experimental data is much better. However, the matching between (iii) ONIOM basis set and experimental structure is the best. This is expected since 6-311++G(d,p) basis set is known to work for organic molecules while LanL2DZ basis set for metal atoms.[35] So, a basis set that combines the two is expected to fare better results.

The optimized geometry generated employing the three basis sets (i), (ii) and (iii) described above are shown in **Figure 4 (a)-(c)** respectively. Comparing the structure obtained from SCXRD (c.f. Figure 2 (a)-(c)), it can be seen that the simplest unit generated by the DFT calculations employing ONIOM basis set match the asymmetric unit obtained by SCXRD very well with some difference in the Cu-O6-C13 angle and Cu-O5-H6 angle. Few of the calculated as well as the experimental geometrical parameters (bond lengths, bond angles and dihedral angles) of **1** are given in **Table 2**. Complete details of the geometrical parameters can be found in the supplementary information. It is to be observed that the bond-lengths as well as the bond angles have an agreement with the experimental values, given the fact that the calculations have been done on a single isolated molecule in the gaseous state while the experimental values correspond to molecules in the solid state.

### 3.2. Rietveld refinement

To ensure that no competing co-crystallizing phases grow using this new technique or phases within **1** grow that are hard to detect in a SCXRD, we resorted to Rietveld refinement of **1**. Red filled circles in **Figure 5** represent the data points obtained in a powder diffractogram that was obtained by crushing few single crystals of **1** and grinding them together for 10 minutes to make a powder. The grinding time of 10 minutes was chosen to ensure that a homogeneous powder with sufficient number of random grains necessary for diffraction is obtained. Details of the calculations are provided in the supplementary. Black solid line is a Rietveld fit that was obtained using the JANA software. [42,43] **Table 3** summaries the obtained parameters of the fit, while **Table 4** describes the fractional co-ordinates of all the atoms, the Debye-Waller factor, $U_{iso}$, as well as occupancies of each atom. The goodness of fit, GoF, is obtained by minimizing the weighted sum of the squared difference between the observed and calculated values of the intensity using the least square methods, and obtained as 3.45. The difference curve plotted as blue curve in Figure 5, shows a small finite value only for few peaks (for most peaks it is negligible) due to intensity mismatch at such positions that may arise due to conformal changes in pyridine or oxalate molecule which were not refined. The small value of GoF was obtained without having to incorporate any preferred orientation function, intergrowth phases or co-crystallizing phases. One can conclude from the SCXRD measurements that our growth technique affords good single crystals and from the Rietveld analysis that there are no competing intergrowth or co-crystallizing phases.

### 3.3. Thermal analysis

To check the thermal stability of the grown crystals, TGA measurements, as shown in **Figure 6**, were done. Figure 6 shows a thermogram of **1** collected in the temperature range from room temperature up to 1000 °C. Red solid line represents the TGA measurement while the blue dotted line shows the DSC measurement done simultaneously to determine the nature of the underlying change (endothermic or exothermic). The weight change in **1** is seen to happen in two steps. In the first step, a shallow peak is observed in the TGA curve at 177 °C accompanied by an endothermic peak in the corresponding DSC curve, which suggests that the shallow peak at 177 °C corresponds to a weight loss which was found to be 5.5% of the total weight**.** This weight loss is attributed to the loss of water whose calculated value is 5.05%. A unique characteristic of $[Cu(C_2O_4)(4\text{-aminopyridine})_2(H_2O)]_n$ is that the crystal undergoes a distinct color change from deep blue to pale blue during dehydration which is reflected in the water loss in the TGA spectrum. After the shallow peak at 177 °C, a sharp peak in the TGA curve accompanied by a sizeable endothermic peak is observed at 199 °C. The sharp peak corresponds to ∼ 68% of weight loss in the coordination polymer. This weight loss may arise due to loss of gases like $CO_2$ arising from the disintegration of oxalate and 4-aminopyridine. The sharp endothermic peak shows that except for the liberation of crystalline water at 177 °C, this crystal is stable up to 199 °C without any phase transition.

### 3.4. FTIR analysis

**Figure 7 (a)** shows the experimentally obtained FTIR spectrum of **1** at room temperature while the FTIR spectra generated using the three bases sets described above are shown in **Figure 7 (b)-(d)** respectively. From the room temperature FTIR spectra of Figure 7 (a), we assigned the main IR features (cm$^{-1}$, KBr pellet), $v$ = 1655 (s; $v_{as}$(O-C=O)), 1626 (vs; $v_{as}$(C=N)), 1603 (s; $v_{as}$(C=C)), 1520 (m), 1458 (m), 1444 (w; $v$(C-N)), 1363 (w), 1330 (m; $v_s$(O-C=O)), 1209 (s; $\delta$(C-H)), 1060 m, 1026 (m; $v_s$(CO)), 827 (m), 779 (m; $\delta$(O-C=O)).[44] By comparing the position and intensity of theoretical and the experimentally obtained spectra, it can be said that the IR frequencies obtained by ONIOM basis set (iii), as shown in Figure 7 (d) match the experimentally obtained frequencies the best. These frequencies are labeled at 1650 (m; $v_{as}$(O-C=O)), 1617 (vs; $v_{as}$(C=N)), 1552 (m; $v_{as}$(C=C)), 1501 (m), 1441 (w; $v$(C-N)), 1360 (m), 1346 (s; $v_s$(O-C=O)), 1210 (m; $\delta$(C-H)), 1063 (m), 1018 (s; $v_s$(CO)), and 840 (w), 777 (w; $\delta$(O-C=O)).

The DSC spectrum of Figure 6 revealed a broad endothermic peak at ~ 640 °C. To understand the reason for this broad endothermic peak, we did FTIR measurements on the title compound both at room temperature as well as at elevated temperatures of 220 °C and 650 °C, as shown in the spectra of **Figure 8 (a)-(c)**, measured in the solid state. In Figure 8 (b) and (c), the main intense peak at 1655 (s; $v_{as}$(O-C=O)) as well as the other lower intensity peaks at lower wavenumbers reduced substantially in intensity after heat treatment at 220 °C and 650 °C respectively. The reduction in intensities of the bands at 1655 (s; $v_{as}$(O-C=O)), 1626 (vs; $v_{as}$(C=N)) and 1603 (s; $v_{as}$(C=C)) indicates the disintegration of the oxalate molecule and aminopyridine groups. The reduction of intensities of bands at 1330 (m; $v_s$(O-C=O)); 1209 (s; $\delta$(C-H)), 1060 (m), 1026 (m; $v_s$(CO)) and 779 (m; $\delta$(O-C=O)) further underscores the degradation of oxalate molecule. So, the broad endothermic peak observable at ~ 650 °C in

Figure 6 corresponds to the near-complete degradation of amino-pyridine groups and oxalate molecule, resulting in the final degradation of the coordination polymer.

### 3.5. Magnetic properties

The temperature-dependent magnetic susceptibility of **1** measured at an applied field of 1 Tesla is shown in **Figure 9**. For the analysis, single crystals of the total mass of 0.975 mg were co-aligned and mounted on the sample holder with vacuum grease. Filled red circles in the main panel of Figure 9 show the magnetic susceptibility of the title compound in the entire temperature range from 300 K up to the lowest measured temperature of 0.49 K. As seen from Figure 9 (a), the magnetic susceptibility increases with decreasing temperatures and passes through a maximum value at ∼ 1.9 K. The presence of a low-temperature maximum is a clear indication of the existence of quasi-one-dimensional antiferromagnetic exchange interactions[28] governed by the Hamiltonian, Equation 1:

$$H = J \sum_{i=1}^{n-1} \vec{S_i} \cdot \vec{S_{i+1}} \qquad (1)$$

where $\vec{S_i}$ represents the spin on the i$^{th}$ magnetic ion, namely, Cu, and the summation is over all nearest neighboring Copper spins. Antiferromagnetic interaction between copper ions along the chain arises due to the overlap between two neighboring magnetic $d_{x^2-y^2}$ orbital of Cu(II) ions resulting in delocalization of spin density on the bridging oxalate ligand.

The black continuous curve in Figure 9 is a fit to the expression:

$$\chi(T) = \chi_0 + \chi_1 + \chi_{BF}(T) \qquad (2)$$

where $\chi_0$ is a small positive constant to account for uncoupled spins; $\chi_1$ corresponds to the diamagnetic contributions from closed atomic shells of the title compound, estimated from Pascal's table[45] and $\chi_{BF}(T)$ is the magnetic susceptibility estimated from the Bonner-Fisher model:[46]

$$\chi = \frac{Ng^2\mu_B^2}{k_BT} \frac{(0.25+0.074795x+0.075235x^2)}{(1.0+0.9931x+0.172135x^2+0.757825x^3)} \qquad (3)$$

where symbols have their usual meaning and $x = J/k_B T$. From the fits, $\chi_o$ and $\chi_1$ were obtained as 4.44 x $10^{-5}$ and -107 x $10^{-6}$ respectively. Reduced chi-square C, defined as C = (1/DOF)*$\sum_i [(\chi_{obs})(i) - (\chi_{cal})(i)]^2$, where DOF is the degree of freedom, was found to be 2.74 x $10^{-9}$. An attempt to fit the data with any possible interchain coupling constant J'/$k_B$ in the mean field approximation resulted in an order of magnitude small values, and hence, was neglected. So, the goodness of fit between the theoretical expressions 2, 3 and the experimental data, points to the fact that the title compound is a very good representation of a spin 1/2 Heisenberg antiferromagnetic chain, with negligibly small interchain coupling constant J', in agreement to other reports.[27,28] The slight difference between the experimental data and that arising from the theoretical fit at low temperatures may be due to the fact that the Bonner-Fisher model calculates the susceptibility of a spin ½ AfHc considering only 11 spins while the experimental data is for very large number of atoms. From the fit, an intra-chain coupling $J/k_B$ = (3.16 ± 0.06) K was obtained, in exact agreement with the value obtained from[27] (2.2 $cm^{-1}$ = 3.16 K). The obtained values of g is 2.27.

## 3. Conclusion

In conclusion, we have successfully synthesized large sized (of length ∼ 1 mm) single crystals of a copper coordination polymer [Cu($C_2O_4$)(4-aminopyridine)$_2$($H_2O$)]$_n$ via a novel and simple growth technique compared to the existing techniques of crystal growth of this compound. Crystals, grown through liquid/liquid or layer diffusion technique, were characterized by different experimental techniques. Density-functional theory with different basis sets was employed to generate the asymmetric unit of the polymer, bond-lengths, etc. and compared with experimentally obtained data. Magnetic susceptibility measurements reveal the compound to be an extremely good representation of a spin 1/2 Heisenberg chain. The largeness of the crystals makes them an attractive candidate for neutron scattering experiments.

**Supporting Information**
Supporting Information is available from the Wiley Online Library or from the author.

**Acknowledgements**
The authors thank M Padmanabhan for his help and support in the initial stages of the synthesis of the crystals.

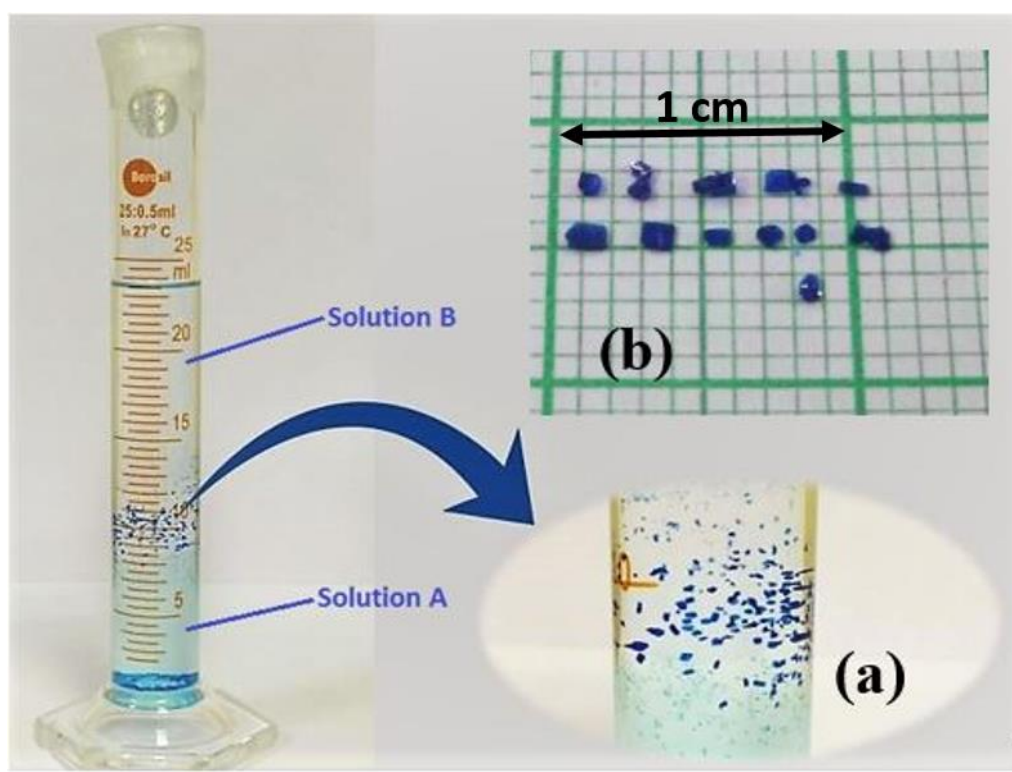

**Figure 1**. Cylindrical vessel showing solution A at the bottom and solution B at the top for synthesizing **1** employing liquid-liquid diffusion. Inset (a) shows the grown crystals at the liquid-liquid interface Inset (b): Images of extracted crystals put on a graph paper for size comparison.

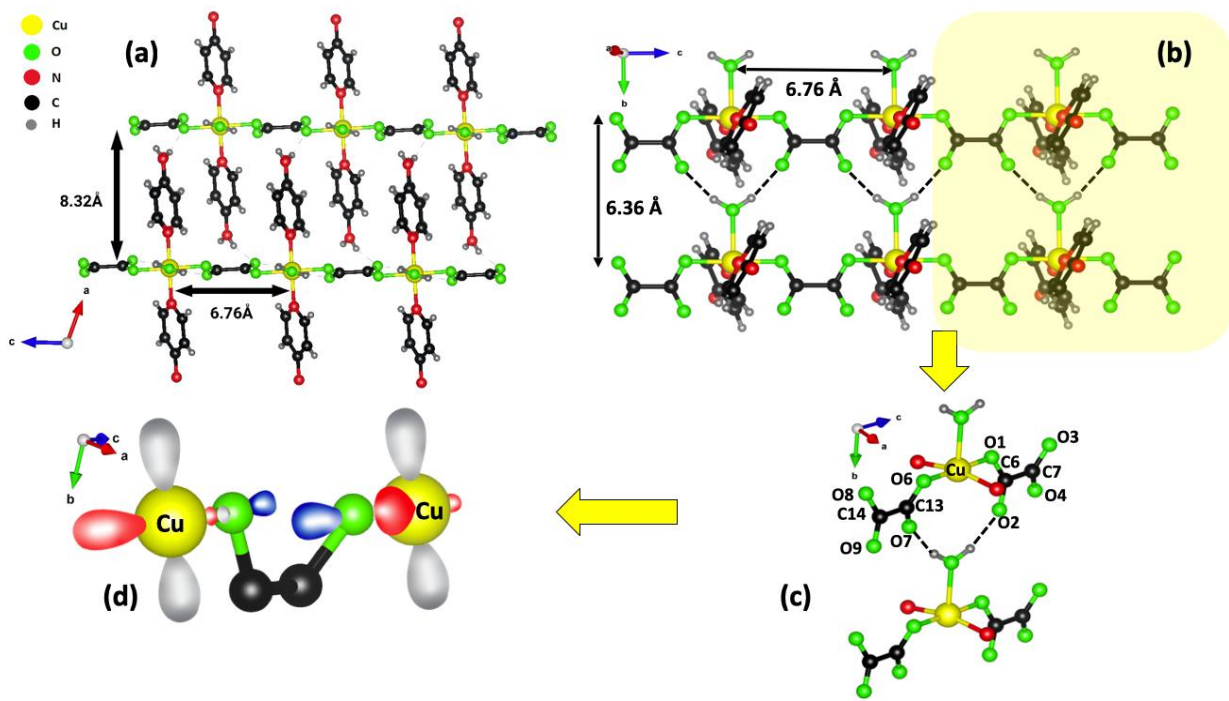

**Figure 2.** Crystal structure of **1**. (a) shows the structure in the a-c plane while (b) shows the structure in the b-c plane. (c) shows the geometry of the oxalate ligand. (d) Schematic representation of magnetic interaction via $d_{x^2-y^2}$ orbital of Cu(II) ions and $p_x$ orbitals of the oxygen atoms of the bridging oxalate ligands.

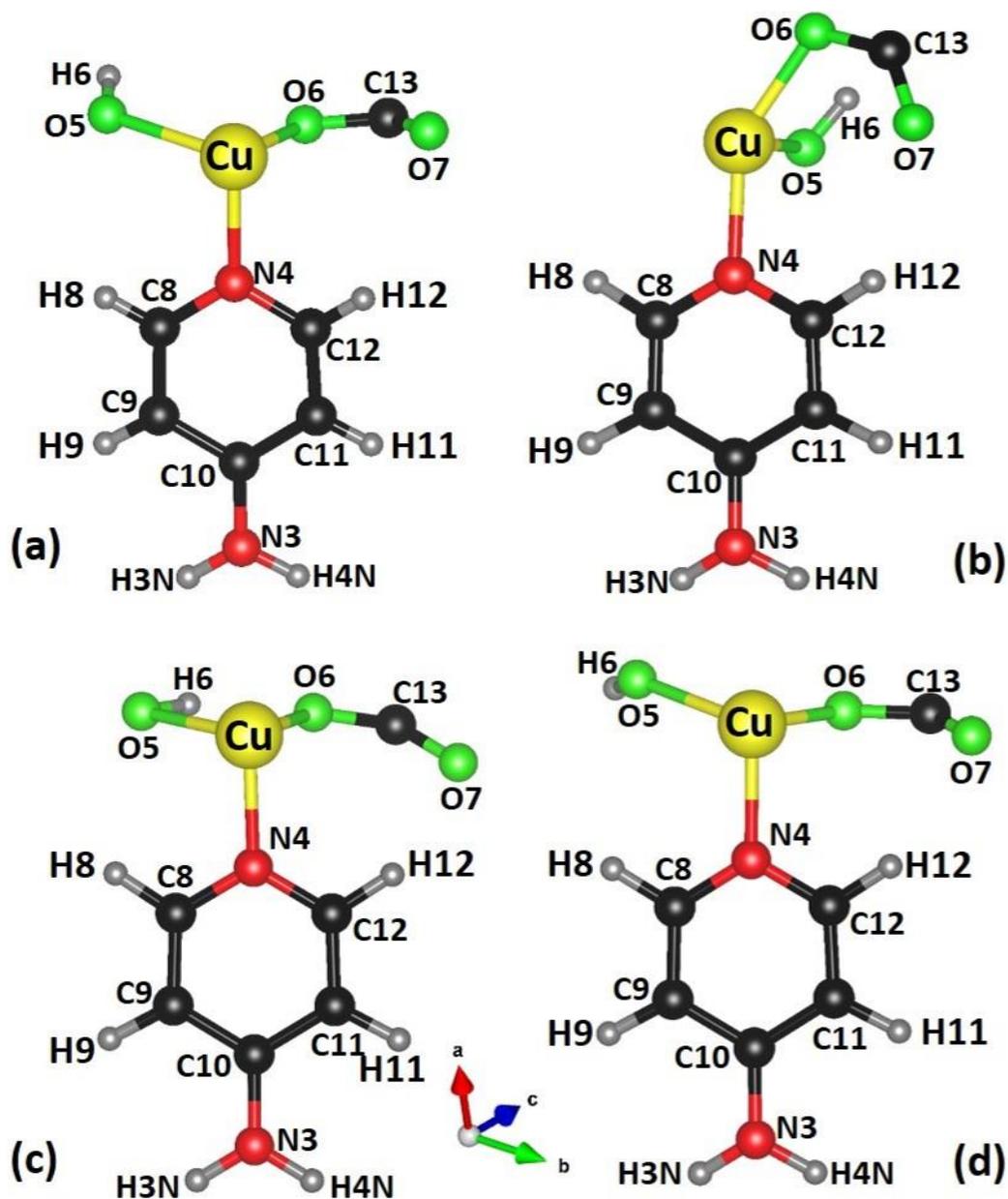

**Figure 3.** (a) Asymmetric unit generated from single crystal XRD. Simplest unit generated from density functional theory using basis sets (b) (B3LYP/6-311++ G(d,p)), (c) (B3LYP/LanL2DZ) and (d) (B3LYP/6-311++ G(d,p), B3LYP/LanL2DZ).

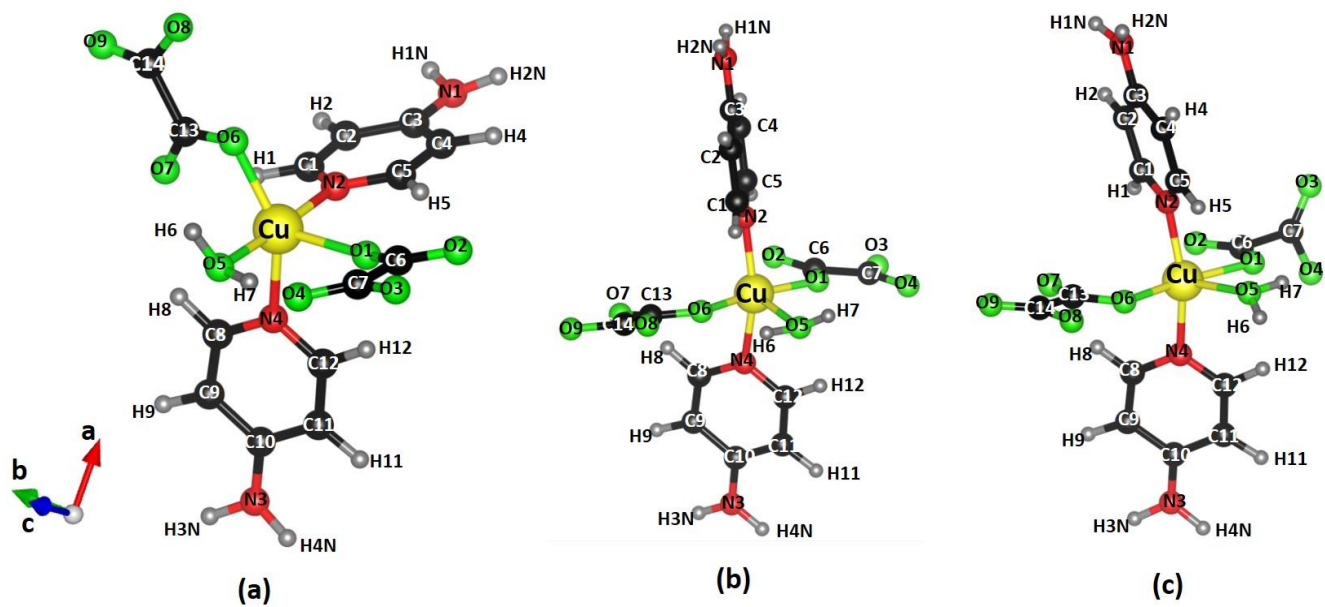

**Figure 4**. Monomeric unit of **1** generated using density functional theory using basis sets (a) (B3LYP/6-311++ G(d,p)), (b) (B3LYP/LanL2DZ) and (c) (B3LYP/6-311++ G(d,p), B3LYP/LanL2DZ).

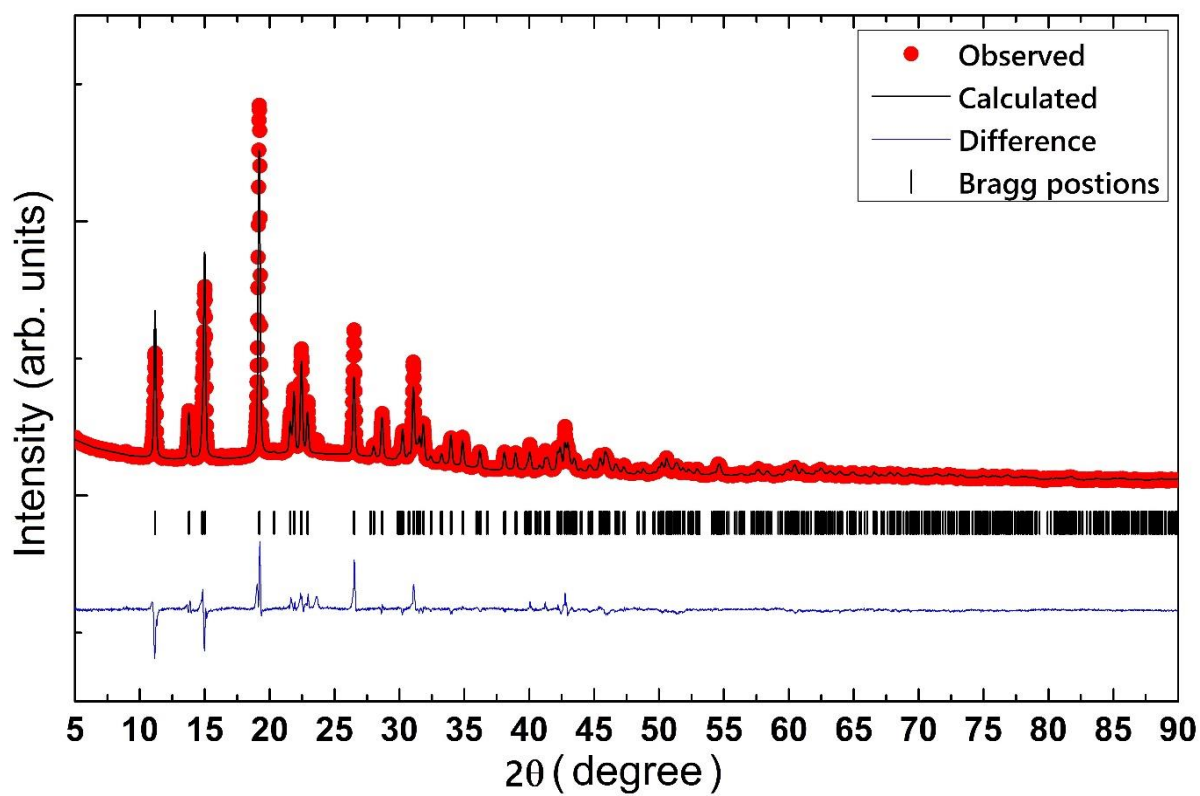

**Figure 5.** Rietveld fit of [Cu(C$_2$O$_4$)(4-aminopyridine)$_2$(H$_2$O)]$_n$ powder where red filled circles indicate the observed data, black solid line is the fit, blue line is the difference curve and black bars indicate Bragg positions corresponding to the [Cu(C$_2$O$_4$)(4-aminopyridine)$_2$(H$_2$O)]$_n$ phase.

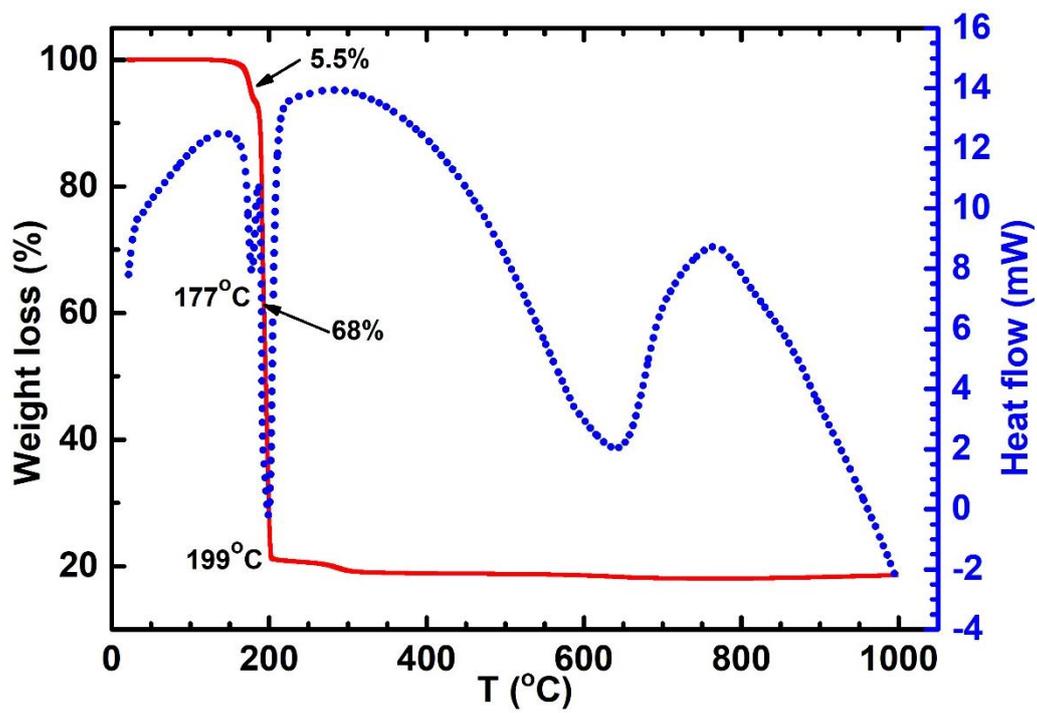

**Figure 6.** TGA (black) and DSC (blue) curves of **1** showing two distinct weight losses.

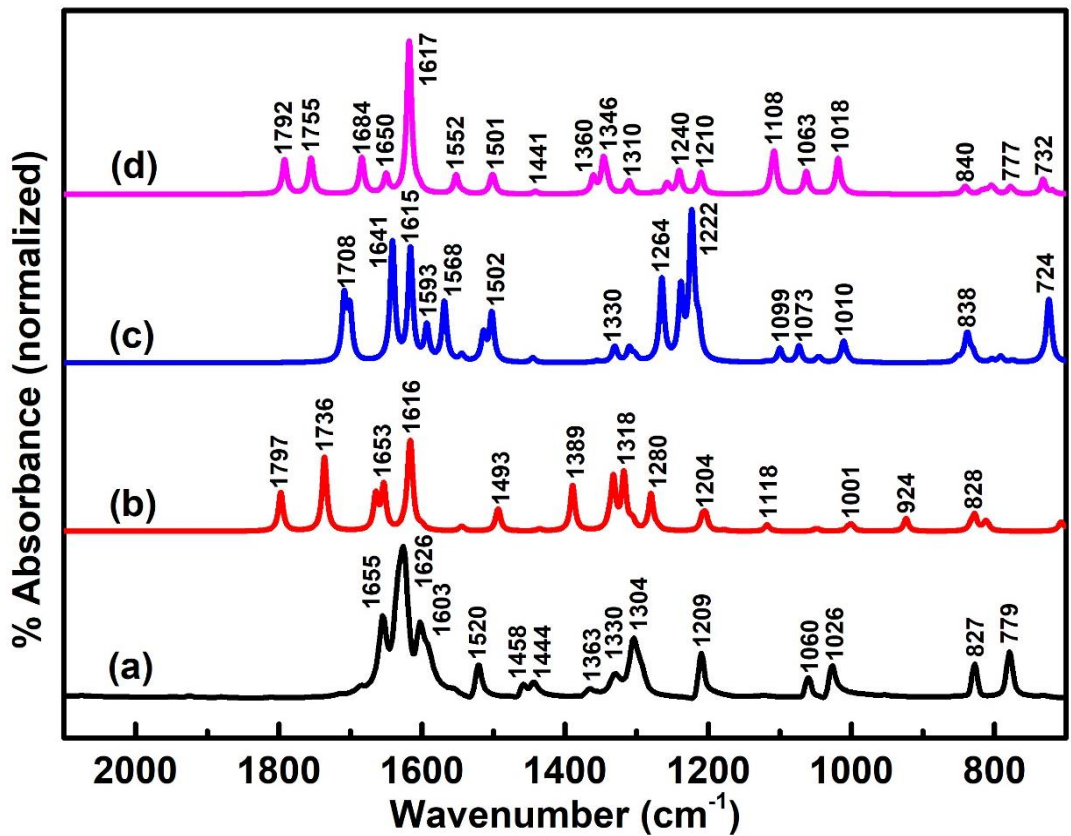

**Figure 7.** (a) Experimental FTIR spectra of **1** at room temperature. FTIR spectra of **1** generated from DFT calculations using basis sets as (b) (B3LYP/6311G++(d,p)), (c) (B3LYP/LanL2DZ) and (d) (B3LYP/LanL2DZ, B3LYP/6-311++ G(d,p).

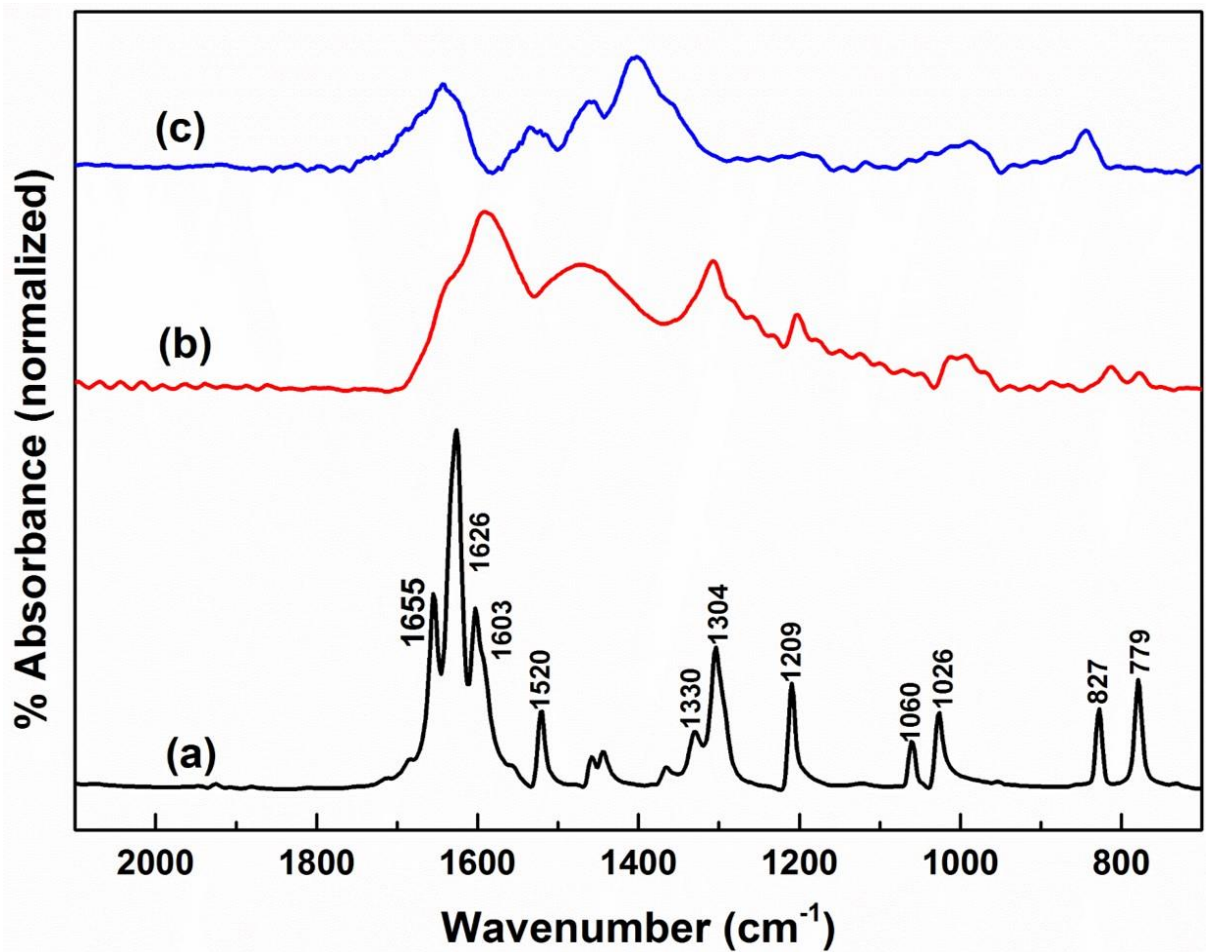

**Figure 8.** FTIR spectra of **1**: (a) as-synthesized, (b) heated at 220 °C and (c) heated at 650 °C.

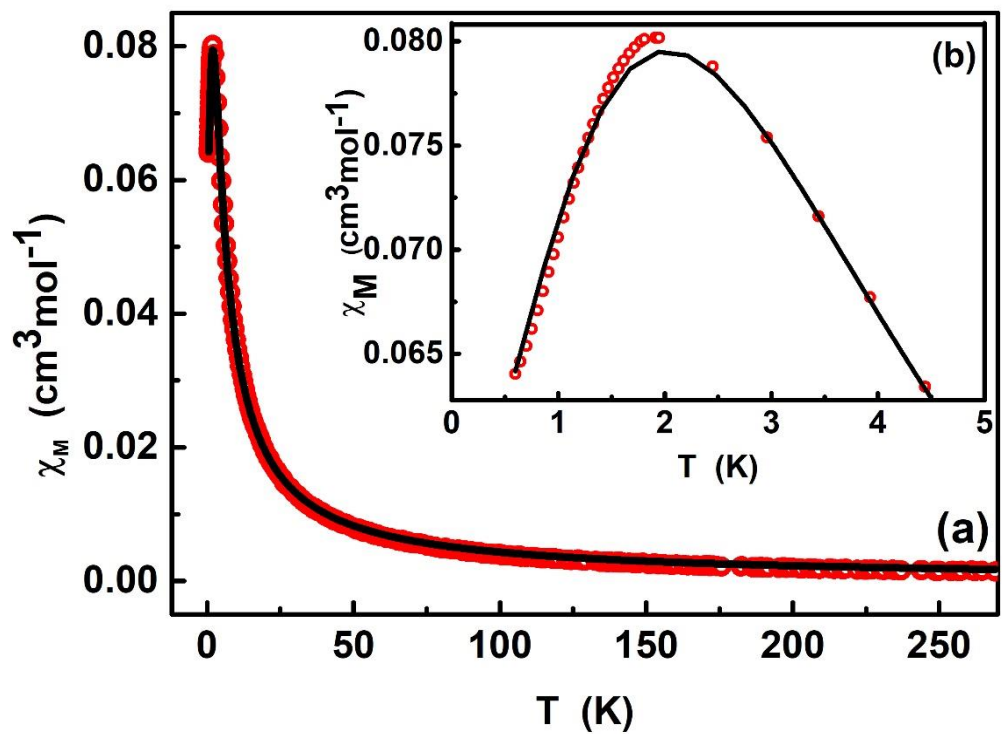

**Figure 9.** Thermal dependence of molar susceptibility, $\chi_M$, of **1** represented by red circles in (a) temperature range of 300 K – 0.49 K and (b) lower temperature range of 0.49 K - 4.2 K. Black solid curve in each is a fit to the Bonner-Fisher expression 2. See text for details.

| | |
|---|---|
| Empirical formula | C12H14CuN4O5 |
| Formula weight | 357.81 |
| Crystal system | Monoclinic |
| Space group | C2 |
| a, Å | 16.6683(18) |
| b, Å | 6.3676(7) |
| c, Å | 6.7632(7) |
| $\beta$, deg | 108.130(3) |
| V, Å$^3$ | 682.19(13) |
| Z | 2 |
| $\rho$ calcd, g/cm3 | 1.742 |
| Goodness-of-fit on F$^2$ | 1.101 |
| Final R indices | R1 = 0.0177, wR2 = 0.0440 |
| R indices (all data) | R1 = 0.0178, wR2 = 0.0440 |

**Table 1.** Crystallographic data of [Cu(C$_2$O$_4$)(4-aminopyridine)$_2$(H$_2$O)]$_n$

| Parameters | Experimental | Calculated (B3LYP/LanL2DZ, B3LYP/6-311++ G(d,p) |
|---|---|---|
| **Bond-Length [Å]** | | |
| Cu-N2 | 1.99833 | 2.02479 |
| Cu-N4 | 1.9983 | 2.02490 |
| **Bond Angle [°]** | | |
| Cu-N2-C5 | 122.7800 | 121.9885 |
| Cu-N2-C1 | 120.3642 | 120.3060 |
| Cu-N4-C12 | 122.7799 | 122.9341 |
| Cu-N4-C8 | 120.3641 | 119.5413 |
| Cu-O6-C13 | 116.8008 | 108.9527 |
| H6-O5-Cu | 122.1022 | 101.4099 |
| **Dihedral Angle [°]** | | |
| Cu-N2-C5-H5 | 2.3260 | 5.9726 |
| Cu-N2-C1-H1 | 2.9925 | 6.8271 |
| Cu-N4-C12-C11 | 177.0 | 177.5867 |
| Cu-N4-C8-C9 | 177.7 | 177.8909 |
| Cu-N4-C12-H12 | 3.0 | 2.0565 |
| Cu-N4-C8-H8 | 2.3 | 0.7873 |

**Table 2.** Experimental and calculated (B3LYP/LanL2DZ, B3LYP/6-311++ G(d,p) geometrical parameters of [Cu(C$_2$O$_4$)(4-aminopyridine)$_2$(H$_2$O)]$_n$

| Lattice constants | a | b | c | α | β | γ |
|---|---|---|---|---|---|---|
| | 16.64891 | 6.363616 | 6.755812 | 90 | 108.1083 | 90 |
| Profile Parameters | U | V | W | Lx | Ly | |
| | 83.03542 | -82.97326 | 43.26922 | 6.271958 | 1.87654 | |
| Zero Correction | 1.03964 | | | | | |
| Asymmetry correction | Method correction by divergence | HpS/L | HmS/L | | | |

|  |  | 0.000219 | 0 |  |  |  |
|---|---|---|---|---|---|---|
| Space group | C2 (Unique axis b) |  |  |  |  |  |
| Background correction | Manual |  |  |  |  |  |
| Agreement factors | GoF 3.45 |  |  |  |  |  |

**Table 3.** *Various parameters of Rietveld refinement obtained on $[Cu(C_2O_4)(4aminopyridine)_2(H_2O)]_n$ using JANA software.*

| Atom type | Label | x | y | z | $U_{iso}$ | Occ |
|---|---|---|---|---|---|---|
| C | C1 | 0.660880 | 0.731900 | 0.579100 | 0.0247859 | 1 |
| H | H1 | 0.626599 | 0.830738 | 0.491159 | 0.03 | 1 |
| C | C2 | 0.745130 | 0.776000 | 0.661100 | 0.0243561 | 1 |
| H | H2 | 0.766453 | 0.902879 | 0.631209 | 0.029 | 1 |
| C | C3 | 0.799110 | 0.628300 | 0.790500 | 0.0213696 | 1 |
| C | C4 | 0.761350 | 0.445900 | 0.835300 | 0.0242 | 1 |
| H | H3 | 0.793883 | 0.345008 | 0.924115 | 0.029 | 1 |
| C | C5 | 0.675460 | 0.416400 | 0.747200 | 0.0251783 | 1 |
| H | H4 | 0.651580 | 0.294171 | 0.779113 | 0.03 | 1 |
| C | C6 | 0.496410 | 0.681100 | 0.112500 | 0.0173652 | 1 |
| N | N1 | 0.624740 | 0.555100 | 0.617800 | 0.0219131 | 1 |
| N | N2 | 0.882820 | 0.661900 | 0.870100 | 0.0293756 | 1 |
| O | O1 | 0.510980 | 0.511000 | 0.212900 | 0.0200682 | 1 |
| O | O2 | 0.477290 | 0.849100 | 0.177400 | 0.0286705 | 1 |
| O | O3 | 0.500000 | 0.155000 | 0.500000 | 0.0335521 | 0.5 |
| Cu | Cu1 | 0.500000 | 0.514630 | 0.500000 | 0.0167034 | 0.5 |
| H | H5 | 0.918900 | 0.556000 | 0.953000 | 0.042 | 1 |
| H | H6 | 0.904800 | 0.789000 | 0.835000 | 0.019 | 1 |
| H | H7 | 0.487000 | 0.077000 | 0.377000 | 0.12 | 1 |

**Table 4.** *Fractional coordinates, $U_{iso}$ and site occupancy for $[Cu(C_2O_4)(4\text{-}aminopyridine)_2(H_2O)]_n$.*